# Compact LumiCal prototype tests for future e+e- colliders

**V. Ghenescu**[a,1,*]

[a] *Institute of Space Science,*
*Atomistilor 409, P.O.Box MG-23, Bucharest-Magurele, RO-077125, Romania*
*E-mail:* ghenescu@spacescience.ro

**on behalf of the FCAL Collaboration*

ABSTRACT: The FCAL collaboration is preparing large-scale prototypes of special calorimeters to be used in the very forward region at future electron-positron colliders for instant luminosity measurement and a precise measurement of integrated luminosity and for assisting beam-tuning. LumiCal is designed as silicon-tungsten sandwich calorimeter with very thin sensor planes to keep the Molière radius small, thus facilitating the measurement of electron showers in the presence of background. Dedicated FE electronics has been developed to match the timing and dynamic range requirements. A partially instrumented prototype was investigated in a 1 to 5 GeV electron beam at the DESY II synchrotron. In the recent beam tests, a multi-plane compact prototype equipped with thin detector planes fully assembled with readout electronics were installed in 1 mm gaps between tungsten plates of one radiation length thickness. High statistics data were used to perform sensor alignment, and to measure the longitudinal and transversal shower development. In addition, Geant4 MC simulations were done and compared to the data.

[1] Corresponding author.

## Contents

## 1. Introduction

The future complementary research facility to the LHC accelerator is a high-energy $e^+e^-$ linear collider [1, 2] that allows exploring in detail the mechanism of electroweak symmetry breaking and opens a new window into physics beyond the Standard Model. The FCAL collaboration is preparing large-scale prototypes of special calorimeters to be used in the very forward region at future electron-positron colliders for instant luminosity measurement and a precise measurement of integrated luminosity and for assisting beam tuning. The forward region sets challenging requirements on several detector parameters, such as detector compactness, radiation hardness or readout ASICs parameters. A schematic layout of the forward region of CLIC/ILC detectors is shown in Figure 1.

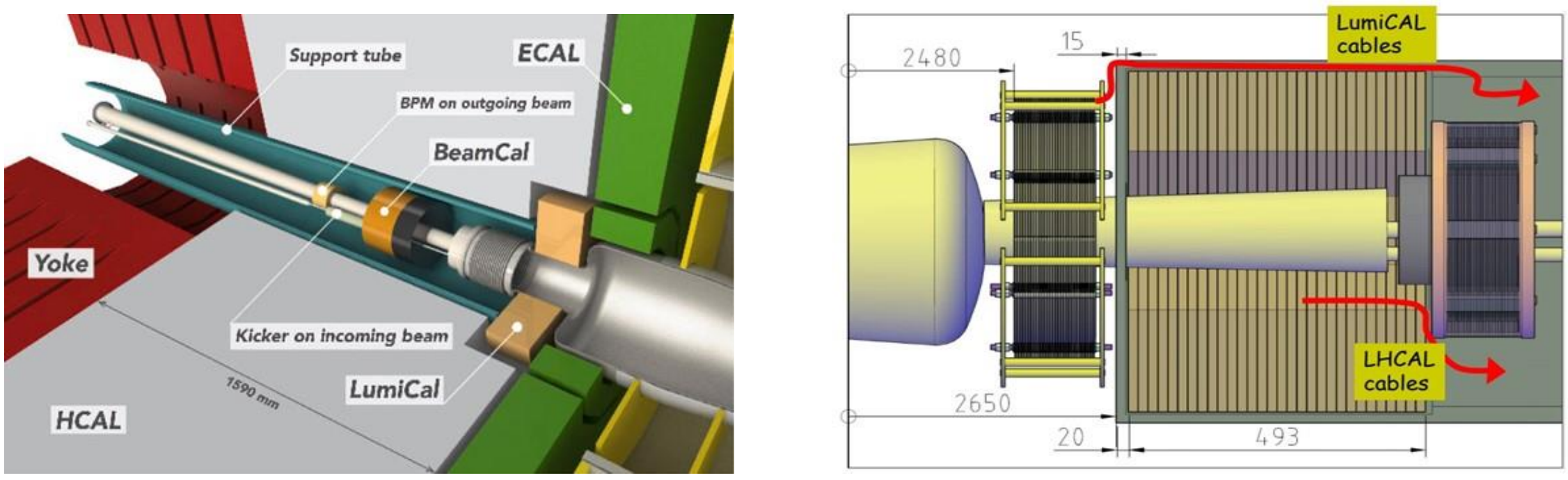

**Figure 1.** The very forward region of the CLIC(left)/ILD(right) detector. LumiCal, BeamCal and LHCAL are carried by the support tube for the final focusing quadrupole and the beam-pipe. TPC, ECAL and LHCAL are the Time Projection Chamber and the Electromagnetic and Low-angle Hadron Calorimeters.

LumiCal and BeamCal are cylindrical electromagnetic calorimeters designed as sensor-tungsten sandwich; the absorber thickness is 3.5 mm corresponding to one-radiation length. The longitudinal structure of both calorimeters consists of 30 absorber layers at ILC (40 at CLIC) interspersed with very thin detector planes. Several solid-state sensor materials [3], including GaAs [4], diamond [5] and single-crystal sapphire [6], along with conventional silicon diode sensors, are being assessed for radiation tolerance. In this paper, the latest developments towards the compact LumiCal prototype calorimeter as well as performance highlights from test-beam campaigns will be presented.

## 2. Thin LumiCal Module

The combination of Si and W for the construction of the detector allows the assembly of a very compact calorimeter made up of active layers with small cell size (high granularity). The sensor is made of a 320 µm thick high resistivity n-type silicon wafer. It has the shape of a sector of a 30° angle, with inner and outer radii of the sensitive area of 80 mm and 195.2 mm, respectively. It comprises four sectors with 64 p-type pads of 1.8 mm pitch. The silicon sensor was glued with epoxy to a fan-out made of flexible Kapton-copper foil. Ultrasonic wire bonding was used to connect conductive traces on the fan-out to the sensor pads. A 70 µm thick Kapton-copper foil glued on the back side of the sensor with a conductive glue supplied the high voltage. For good mechanical stability, the assembled sensor was embedded in a carbon fibre envelope. The fully assembled detector plane, developed for this study, has a total thickness of about 650 µm. A picture of the thin LumiCal module is shown in Figure 2. The detector plane is attached to the tungsten plate by adhesive tape. The absorber tungsten plate is glued in a permaglass frame which could be fixed in one of the 30 comb slots of the mechanical framework. Tungsten plates are made from an alloy of 93% tungsten, 5% nickel and 2% copper. The flatness [7] of tungsten plates is measured to be better than 30 µm. The flatness map of a 140x140x3.5 $mm^3$ tungsten plate is shown in Figure 3. Flatness measurements have been done at predetermined points (X and Y) on the surface of each tungsten plate.

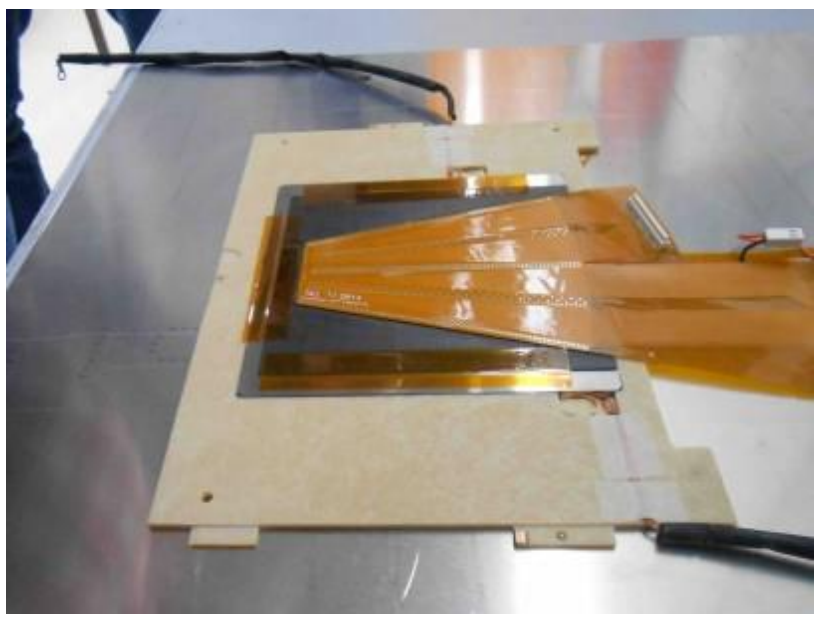

**Figure 2.** Photograph of a thin LumiCal module assembly glued on the tungsten plate.

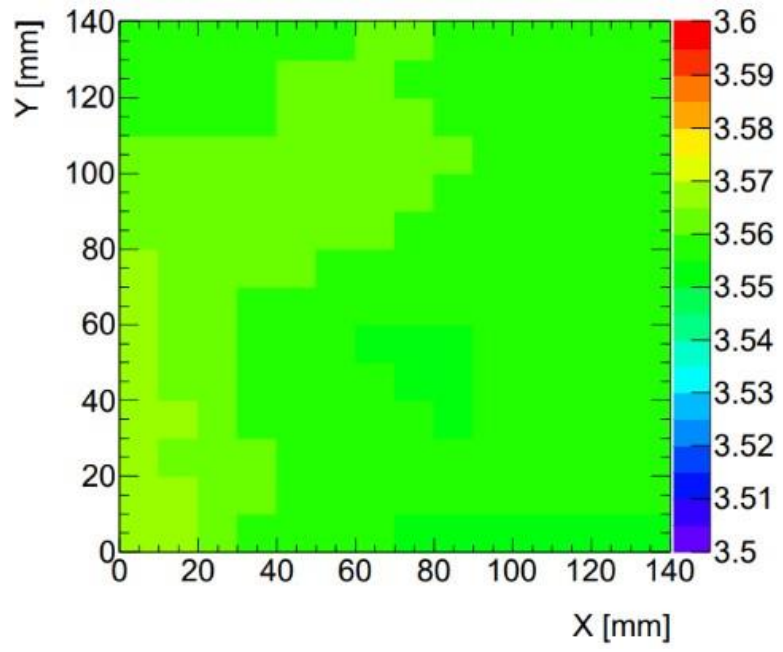

**Figure 3.** Flatness map of the tungsten plate.

## 3. New readout ASIC

Until the 2020 test-beam campaign at the DESY II Synchrotron the FCAL Collaboration used the APV-25 chip hybrid board as a temporary solution. The APV-25 chip was designed for the CMS track detectors and it is optimized for reading signals from calorimeter. Last year three new readout boards based on new readout ASIC called FLAME [8] (FcaL Asic for Multiplane rEadout) were manufactured and were connected to the LumiCal detectors to record the channel signals. This FLAME was designed in 130 nm CMOS technology and matches the requirements of a future linear collider, in particular it has a larger input signal dynamic range compared to the APV-25 chip. It consists of a pair of two identical 16-channel blocks. A block diagram of FLAME with 32-channel ASIC is shown in Figure 4. The main features of the new readout board are summarized in table 1.

## 4. Test-beam results

In recent years the FCAL Collaboration has carried out several experiments to study the performance of the ultra compact LumiCal calorimeter. The events collected at the DESY-II using electrons with energies between 1 GeV to 5 GeV have been used to measure the precision of the shower position determination, the electromagnetic shower development in longitudinal and transverse directions and the effective Molière radius. The measurement results from the test beam were compared with prediction of GEANT4 Monte Carlo simulations where the experimental setup was implemented. The most recent test beam campaign took place in 2020 at DESY II and tested first deep LumiCal prototype calorimeter with 15 detector planes installed in 1 mm gap between the tungsten absorber plates. A sketch of the experimental setup is shown in Figure 5 (left) and a and a photo of the assembled LumiCal stack with the first 3 planes connected to the FLAME readout board (right). These three FLAME boards were connected in turn to all the sensors in the LumiCal stack. Figure 6 shows the preliminary results of signal distribution produced by a 5 GeV electron beam and measured in a single channel of the first detector plane. The most probable value (MPV) of the peak is estimated using a fit with Landau–Gauss convolution function. A map with the MPV values of the amplitude distribution for each channel is shown in Figure 7. Figure 8 shows preliminary results of the calorimeter response to 5 GeV electron beam readout with the FLAME ASICs. Using the same approach like in Ref. [9] the effective Molière radius will be estimate and the results will be compared with the MC simulation.

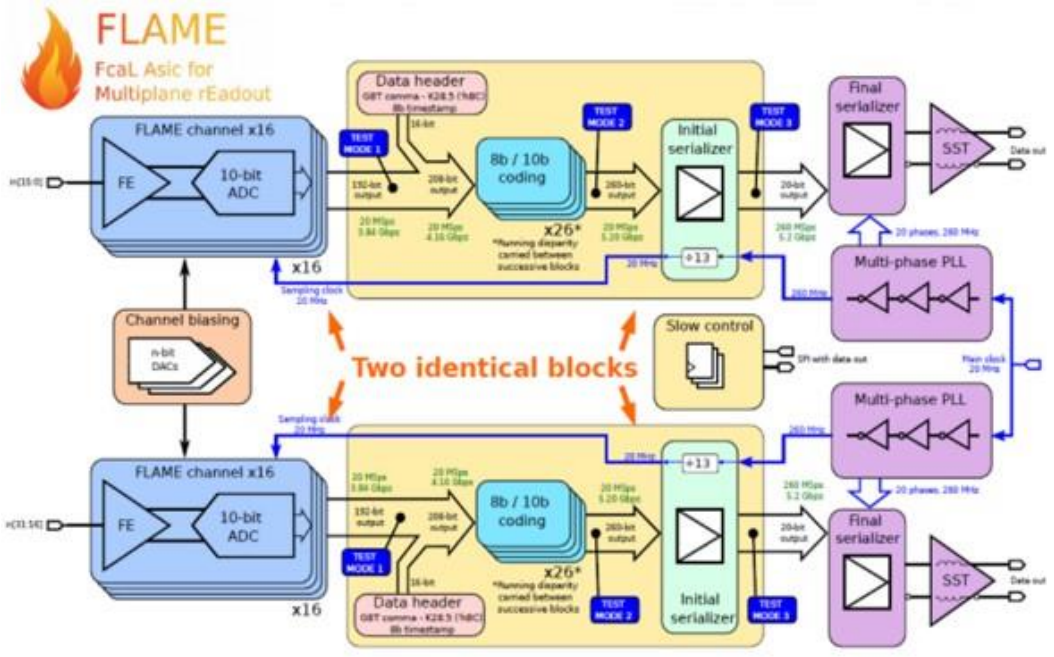

**Figure 4.** Block diagram of 32 – channels ASIC FLAME architecture.

| | **Features** |
|---|---|
| Process Technology | 130 nm TSMC CMOS |
| Channels per ASIC | 32 |
| Chip dimension | 3.7 mm x 4.3 mm |
| Power dissipation | ~ 2 mW/ch |
| ADC digitization | 10-bits |
| Sampling rate | Up to 20 MS/s |
| Output rate | 5.2 Gb/s |

**Table 1.** FLAME ASIC – main specifications.

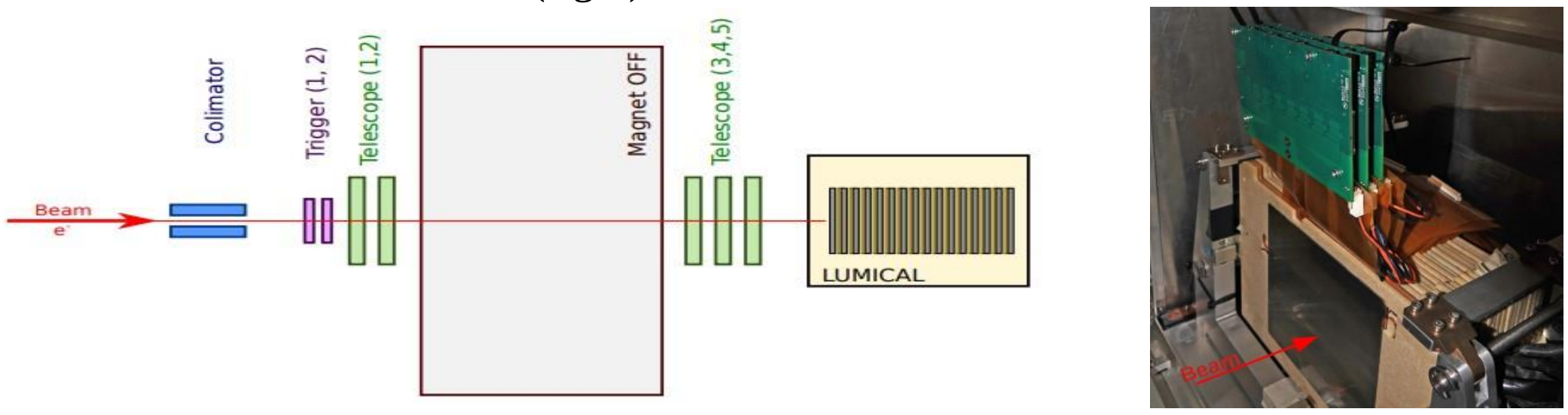

**Figure 5.** (left) Schematic view of the experimental set-up (not to scale). (right) A close-up on the LumiCal detectors, three of them connected to the FLAME readout.

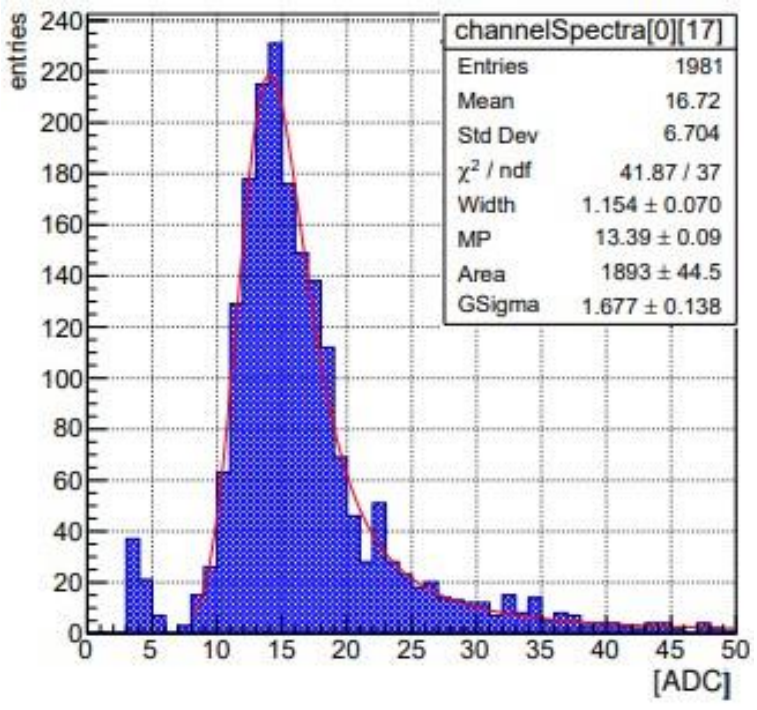

**Figure 6**. Signal distribution in a single channel fitted with LG convolution function.

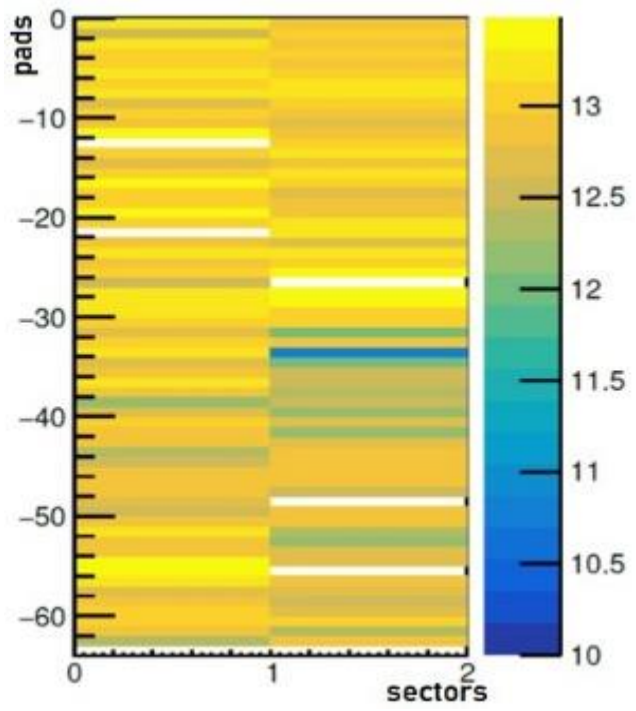

**Figure 7**. MPV value of the amplitude distribution map for each channel.

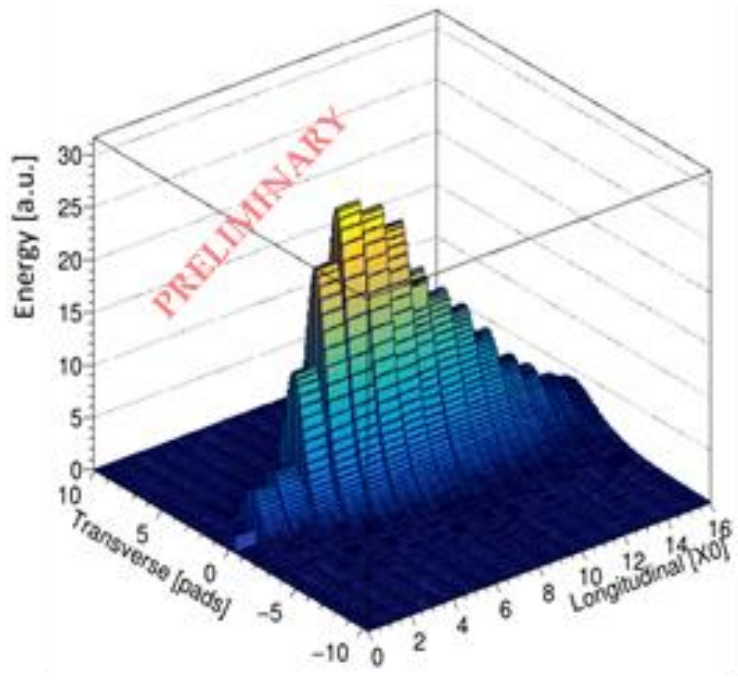

**Figure 8**. A lego plot of the transvers profile for each layer from the beam-test data.

## 5. Conclusions

Major components developed by FCAL Collaboration can be operated as a system in the future linear collider. Thin detector module with submillimeter thickness has been developed and produced. Its geometry matches requirements of the LumiCal conceptual design. The calorimeter prototype with 15 sensitive layers were tested in electron beam and the data analysis in progress. A new electronic readout board called FLAME has been produced and tested in an test-beam campaign.

The performances of the LumiCal detector demonstrate the feasibility of constructing a compact calorimeter consistent with the conceptual design, which is optimized for a high precision luminosity measurement in e+e- collider. Technologies developed in FCAL Collaboration are applied in other experiments like CMS, XFEL and are considered for LUXE at DESY.

## Acknowledgments

This activity was partially supported by the Romanian UEFISCDI agency under grant no. 16N/2019. The measurements leading to these results have been performed at the Test Beam Facility at DESY Hamburg (Germany), a member of the Helmholtz Association (HGF).

## References

[1] T. Behnke et al., The International Linear Collider. Technical Design Report, Volume 4: Detectors, 2013. arXiv:1306.6329 [physics.ins-det].

[2] The CLIC, CLICdp collaborations, Updated baseline for a staged Compact Linear Collider. CERN Yellow Report CERN-2016-004; arXiv:1608.07537 [physics.acc-ph].

[3] P. Anderson et al., Updated Results of a Solid-State Sensor Irradiation Study for ILC Extreme Forward Calorimetry, Dec., 5–9, 2016, LCWS, Morioka, Japan, https://arxiv.org/pdf/1703.05429.pdf

[4] K. Afanaciev et al., Investigation of the radiation hardness of GaAs sensors in an electron beam, 2012 JINST 7 P11022.

[5] Ch. Grah et al., Polycrystalline CVD Diamonds for the Beam Calorimeter of the ILC, IEEE Trans. Nucl. Sci. 56 (2009) 462.

[6] O. Karacheban et. al, Investigation of a direction sensitive sapphire detector stack at the 5 GeV electron beam at DESY-II, 2015 JINST 10 P08008.

[7] Mikhail Gostkin, Tungsten plates, 32nd FCAL Collaboration Workshop, May 10 – 11, 2018, IFJ PAN Cracow, https://indico.cern.ch/event/697164/timetable/#20180510.detailed .

[8] Jakub Moron et al., FLAME, a new readout ASIC for the LumiCal detector, CLIC Workshop 2016, CERN, https://indico.cern.ch/event/449801/contributions/1945306/

[9] H. Abramowicz, et al., Measurement of shower development and its Molière radius with a four-plane LumiCal test set-up, Eur. Phys. J. C 78 (2018) 135